\documentclass[9pt,twocolumn, notitlepage]{article}

\usepackage[backend=biber,style=nature,sorting=none,maxbibnames=99]{biblatex}
\usepackage{amssymb}
\usepackage{amsmath}

\usepackage{graphicx}

\usepackage{setspace}
\usepackage[top = 1in, bottom = 1in, left = .75in, right = .75in]{geometry}
\graphicspath{ {./images/} }
\newcommand{\Mgeo}{M_{geo}}
\usepackage[labelfont=bf,format=plain]{caption}

\renewcommand{\abstractname}{\vspace{-\baselineskip}}

\renewenvironment{abstract}{%
  \small
  \begin{center}%
    {\bfseries \abstractname\vspace{-0.5em}\vspace{0pt}}%
  \end{center}%
  \quotation}
  {\endquotation}

\usepackage{xcolor}

\begin{document}

\title{Superfluid response of an atomically thin, gate-tuned van der Waals superconductor}
\author{Alexander Jarjour$^1$, G.M. Ferguson$^1$, Brian T. Schaefer$^1$, Menyoung Lee$^{2,3}$, \\ Yen Lee Loh$^4$, Nandini Trivedi$^5$, Katja C. Nowack$^{1,2,0}$}
\date{}
\twocolumn[
\maketitle 
    \begin{abstract}   
        \textbf{A growing number of two-dimensional superconductors are  being discovered in the family of layered van der Waals (vdW) materials. Due to small sample volume, their characterization has been largely limited to electrical transport measurements. As a consequence, characterization of the diamagnetic response of the superfluid to an applied magnetic field, a defining property of any superconductor, has been lacking. Here, we use a local magnetic probe to directly measure the superfluid response of the tunable, gate-induced superconducting state in MoS$_2$. We find that the backgate changes the superconducting transition temperature non-monotonically whereas the superfluid stiffness at low temperature and the normal state conductivity monotonically increase with backgate voltage. In some devices, we find direct signatures in agreement with a Berezinskii-Kosterlitz-Thouless transition, whereas in others we find a broadened, shallow onset of the superfluid response. We show that the observed behavior is consistent with disorder playing an important role in determining the superconducting properties in superconducting MoS$_2$. Our work demonstrates that magnetic property measurements are within reach for vdW superconductors and reveals that the superfluid response significantly deviates from simple BCS-like behavior. }
        
    \end{abstract}
\vspace{\baselineskip}
]
\footnotetext[1]{Laboratory of Atomic and Solid State Physics, Cornell University, Ithaca, NY}
\footnotetext[2]{Kavli Institute at Cornell for Nanoscale Science, Ithaca, NY}
\footnotetext[3]{Department of Electrical and Computer Engineering, Cornell University, Ithaca, NY}
\footnotetext[4]{Department of Physics and Astrophysics, University of North Dakota, Grand Forks, ND}
\footnotetext[5]{Department of Physics, The Ohio State University, Columbus, OH}
\footnotetext[0]{Corresponding author, email: kcn34@cornell.edu}

The two defining properties of a superconductor are a vanishing electrical resistance and the expulsion of magnetic fields below a characteristic critical temperature, $T_c$. Typically, superconductivity is first identified in a material by observing a sharp drop in the resistance at $T_c$. However, resistance measurements only give limited information about the superconducting state forming below $T_c$, and other experimental probes are needed to reveal its nature. Measurements of the strength with which the superconductor screens a magnetic field directly probe the superfluid stiffness, $\rho_s$, and have provided insight into the nature of unconventional superconductors \cite{Prozorov2006, Uemura1989, Lemberger2011,Bozovic2016}. From $\rho_s$, the superfluid density can be extracted, which in a clean BCS superconductor at $T = 0$ is expected to be equal to the normal carrier density \cite{Leggett1998}. Comparing the superfluid density to the normal carrier density, pair breaking by impurity scattering and other mechanisms can be identified \cite{tinkham2004}. In two-dimensional superconductors, the onset of $\rho_s$ may show fingerprints of the Berezinskii-Kosterlitz-Thouless (BKT) transition \cite{Kosterlitz1973, B.I.Halperin1979}. 

A growing family of atomically thin superconductors is realized by mechanically exfoliated sheets of vdW materials. These include two-dimensional (2D) superconductors based on bulk superconducting materials such as NbSe\(_{2}\) \cite{Frindt1972, Staley2009}, NbS$_2$ \cite{Yan2019}, and TaS$_2$ \cite{Navarro-Moratalla2016}, as well as 2D superconductors that are induced by electrostatic gating such as  MoS\(_{2}\) \cite{Iwasa2012Mos2}, WS\(_{2}\) \cite{Jo2015, Zheliuk2017}, MoTe\(_{2}\) \cite{Rhodes2021}, WTe\(_{2}\) \cite{Fatemi2018,Sajadi2018}, twisted bilayer graphene \cite{Cao2018}, and ABC stacked trilayer graphene \cite{Chen2019}. A variety of superconducting phenomena have been observed in atomically thin vdW superconductors, such as robustness against large in-plane magnetic fields \cite{J.M.Lu2015,Xi2016, DeLaBarrera2018}, superconductivity in the vicinity of correlated electronic states \cite{Cao2018, Chen2019}, a dramatically enhanced $T_c$ in the monolayer limit \cite{Rhodes2021, Navarro-Moratalla2016}, and unusual symmetry breaking in the superconducting state \cite{Hamill2021}. A detailed study of the transport properties of NbSe$_2$ with varying thickness has shown that dissipationless transport is highly fragile to temperature, applied magnetic field and the employed bias current \cite{Benyamini2019} further highlighting the need for directly probing the phase coherence of the superconducting state in vdW materials. However, due to the typically small sample size, only a few measurements beyond electronic transport which directly probe the superconducting state below $T_c$ are available \cite{Costanzo2018, Oh2021}, and no characterization of the magnetic response has been reported for any atomically thin vdW superconductor.
\begin{figure}[ht!]
    \includegraphics[width=\columnwidth]{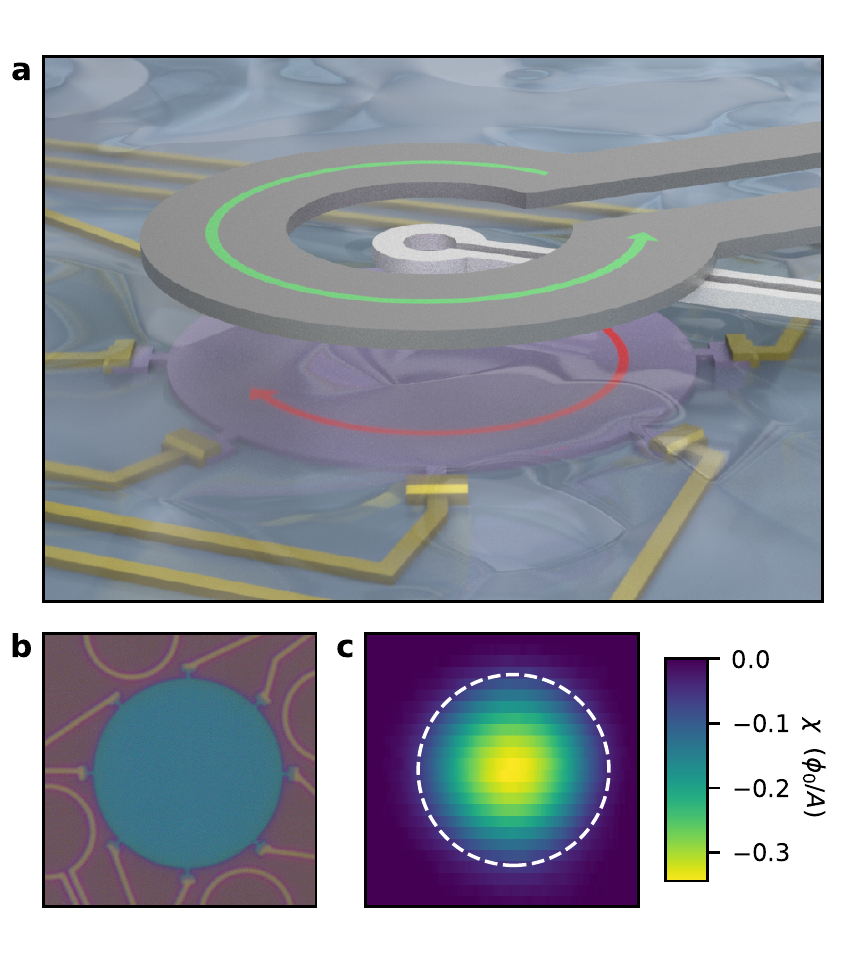}
    \centering
        \caption{ \textbf{Magnetic measurement of ionic gated MoS$_2$}. \textbf{a}, A flake of MoS\(_2\) (shown in purple) on a SiO\(_2\)/Si substrate is patterned into a disk and covered by a spin coated ionic gel. The SQUID pickup loop (shown in silver), with concentric field coil (shown in dark gray) is approached to the sample. A current in the field coil produces a magnetic field, which results in an opposing screening current in the superconductor. The strength of the screening current is magnetically detected by the pickup loop. \textbf{b}, Optical image of a 20 $\mu$m diameter circular MoS\(_2\) device (blue) with electrical contacts. \textbf{c}, Image of the magnetic response of the device shown in \textbf{b} at 4 K. The white dashed circle has a diameter of 20 \(\mu \)m and indicates the device circumference.}
        \label{fig:diagram}
\end{figure}

Here, we report direct measurements of the magnetic response of the gate-induced superconducting state in few-layer MoS$_2$. Although MoS$_2$ is a semiconductor when undoped, ionic liquid gating can induce an electron accumulation layer at the surface of a MoS$_2$ flake which exhibits superconductivity at carrier densities exceeding $0.5 \times 10^{14}$ cm$^{-2}$ \cite{Iwasa2012Mos2}. $T_c$ changes non-monotonically with the carrier density with a maximum of approximately 10 K. Superconductivity is retained in the monolayer limit \cite{Costanzo2016}, but is always in the 2D limit regardless of the flake thickness because the accumulation layer is approximately confined to the topmost layer \cite{Saito2016, J.M.Lu2015}. Spin-valley locking in the electronic bandstructure of MoS$_2$ gives rise to an Ising protection of the superconducting state leading to an in-plane critical field dramatically exceeding the Pauli limit \cite{J.M.Lu2015,Saito2016}. Recently, tunneling measurements have suggested that the order parameter is not fully gapped \cite{Costanzo2018}, a possible signature of an unconventional superconducting state.
 
We fabricate our devices from exfoliated MoS$_2$ flakes with a thickness of 3-10 layers, and pattern them into disks with Ti/Au contacts to allow gating and electrical transport measurements (Fig. \ref{fig:diagram}b). To induce high carrier densities, we use a spin coated, approximately 2-$\mu$m-thick ionic gel. In addition, we apply a backgate voltage, $V_\text{BG}$, across 300 nm thick SiO$_2$ to our devices. Figure \ref{fig:diagram}a schematically shows how we measure the magnetic response of a device. A scanning superconducting quantum interference device (SQUID) \cite{Huber2008,Kirtley2016} with a pickup loop and a concentric field coil is centered above the device. Here the pickup loop and field coil have an inner diameter of 1.5 \(\mu\)m and  8 \(\mu\)m, respectively.  An AC current in the field coil produces a small magnetic field, and we measure the resulting flux in the pickup loop using a lock-in amplifier. Away from the sample, this signal corresponds to the mutual inductance between the pickup loop and field coil. When the pickup loop/field coil pair is brought close to the device, the magnetic response of the device appears as a change in the mutual inductance. By measuring this change, we directly probe the device's magnetic response \(\chi\).

Specifically, a superconductor generates currents to screen the applied magnetic field. The strength of the screening currents can be related to the superfluid stiffness $\rho_s$, the Pearl length $\Lambda$, and the superfluid density $n_s$ which are connected through \(\rho_s = \hbar^2/(2 \mu_0 k_B e^2 \Lambda)\) and \(n_s^{2D}/m^* = 4 k_B \rho_s/\hbar^2 \), where $\hbar$ is the reduced Planck's constant, $\mu_0$ is the permeability of free space, $k_B$ is the Boltzmann constant, $e$ is the elementary charge, and $m^*$ is the effective mass. In the limit of weak screening, i.e., $\Lambda \gtrsim R$, where \textit{R} is the device size, the magnetic response \(\chi\) is directly proportional to $\rho_s$: 
$\chi = \Mgeo \rho_s$, which is why we refer to $\chi$ in the following as the superfluid response. The proportionality factor $\Mgeo$ depends on the SQUID dimensions, its height above the sample, and the device dimensions. We can model $\Mgeo$ to extract absolute values of $\rho_s$ and $\Lambda$ from our measurements (Supplementary Information). The estimate of $\Mgeo$ has systematic uncertainties due to uncertainties in the SQUID height, the exact device dimensions, and other geometrical factors; however, relative changes of $\rho_s$ as a function of $V_\text{BG}$ and temperature are captured with high accuracy. Figure \ref{fig:diagram}c shows an image of  \(\chi\) at a constant height. From this, we identify the center of the device where we position the SQUID. We then  measure \(\chi\) at a fixed height as a function of temperature and \(V_\text{BG}\) and simultaneously record the sheet resistance $R_{\square}$.

\begin{figure*}[t!]
    \centering
    \includegraphics[width=\textwidth]{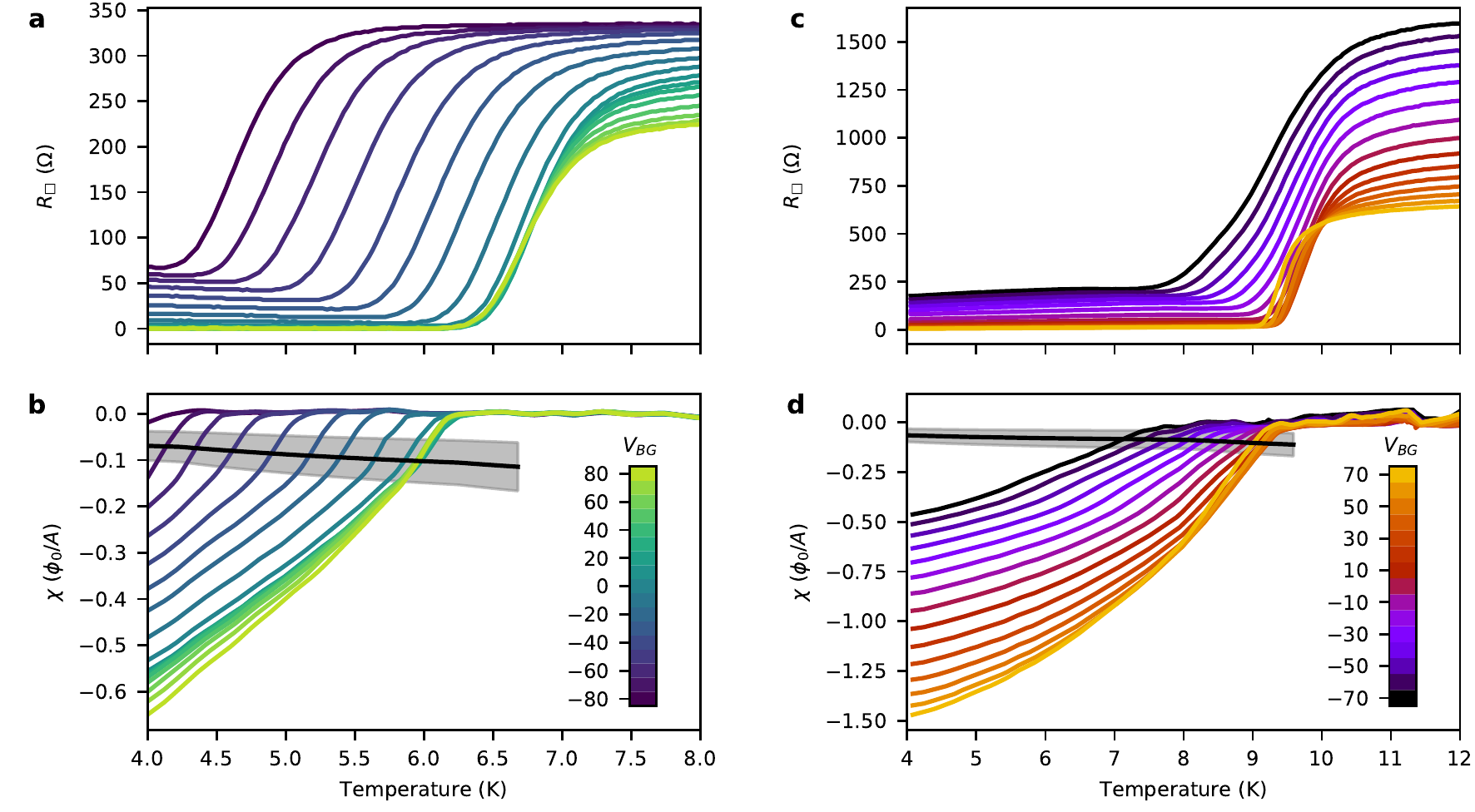}
    \caption{\textbf{Temperature dependence of the gate-tuned resistance and superfluid response. a}, Sheet resistance, $R_{\square}$, and \textbf{b} superfluid response, $\chi$, of device A versus temperature. Different colors correspond to different backgate voltage $V_\text{BG}$ as indicated by the color bar in \textbf{b}. The black line corresponds to the  universal BKT condition $\rho_s = 2T/\pi $. The gray shaded area indicates uncertainty in the universal condition arising from uncertainty in the SQUID height. \textbf{c,d}, Same as in \textbf{a,b} but for device B.}
    \label{fig:bothdev}
\end{figure*}

Figure \ref{fig:bothdev} shows $R_{\square}$ and $\chi$ as a function of temperature and $V_\text{BG}$ for two devices, which we label A (20 $\mu$m diameter) and B (15 $\mu$m diameter). We include data from an additional device in the Supplementary Information (Fig. S3). A weakly temperature-dependent magnetic response from the ionic liquid has been subtracted from $\chi$ (Supplementary Information). For both devices, $R_{\square}$ drops sharply as we lower the temperature, and the superfluid response appears when the drop in resistance is completed. At $V_\text{BG}=0$, device A has a lower $T_c$ of approximately 6 K, compared to 9 K in device B. For both devices, the critical temperature changes with \(V_\text{BG}\). In device A, the resistive transition remains approximately 1.5 K wide across the backgate range, whereas in device B the transition broadens with decreasing $V_\text{BG}$ and changes shape. Likewise, we observe that the shape of the superfluid response versus temperature for device A is qualitatively independent of $V_\text{BG}$, but does vary for device B. However, in both devices, the response measured at the lowest temperature increases monotonically with $V_\text{BG}$. Finally, the resistance below the transition is finite for both devices at low values of \(V_\text{BG}\). Combined with the simultaneous superfluid response of the sample, this raises the question if this behavior is intrinsic to the superconducting state. A possible extrinsic explanation is a non-superconducting region along the periphery of the device, which would have a particularly pronounced effect in disk-shaped devices (Supplementary Information). While determining the origin of the finite resistance is interesting, it is outside the scope of this work. In the following, we first focus on the superfluid response substantially below  $T_c$. We then discuss the onset of diamagnetism close to the superconducting transition.

\begin{figure*}[t!]
    \centering
    \includegraphics[width=\textwidth]{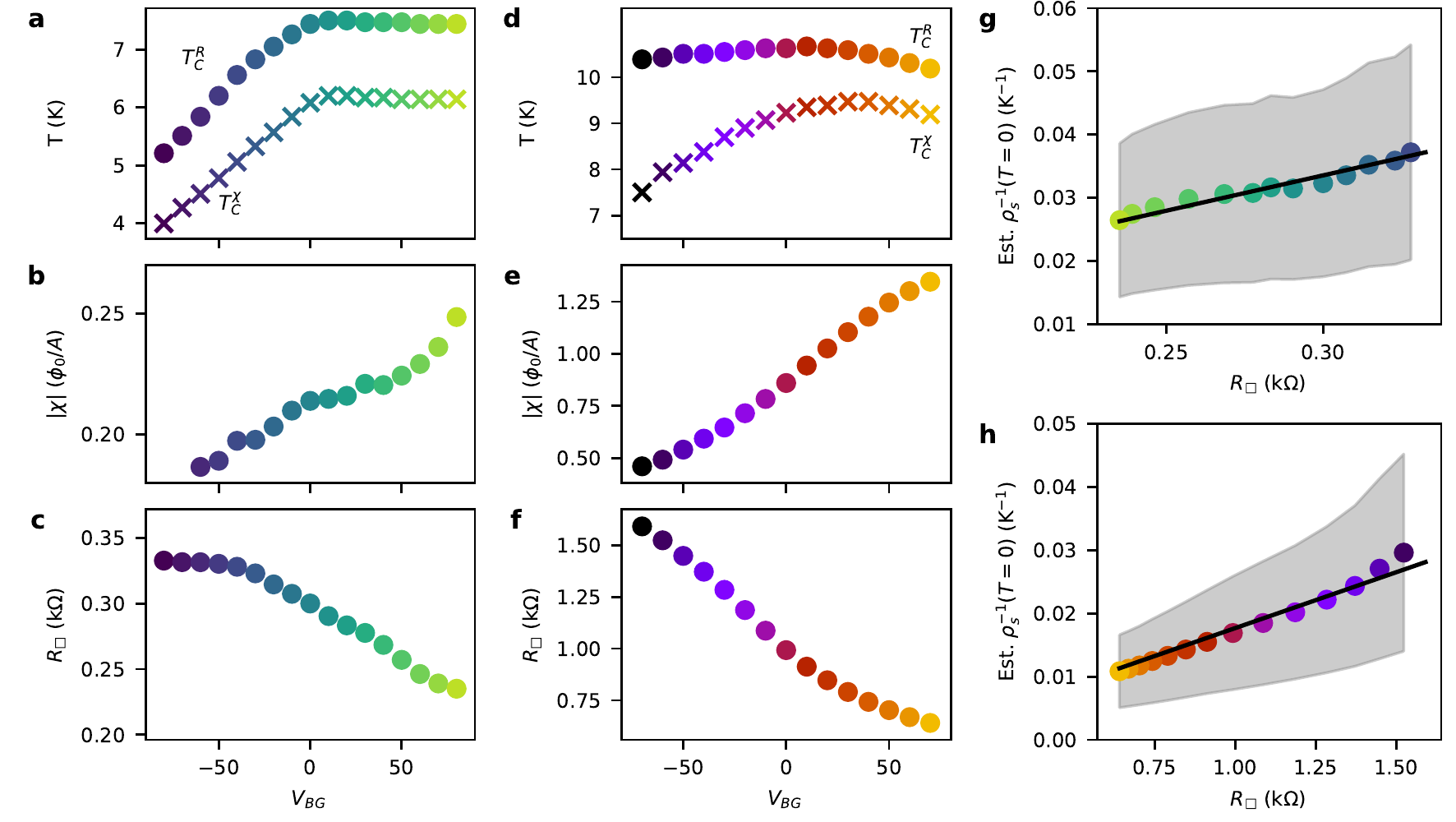}
    \caption{\textbf{Correlation between device resistivity and superfluid stiffness. a-c}, Summary of properties of the superconducting state in device A versus $V_\text{BG}$. \textbf{a}, Two measures of the critical temperature: $T_c^R$ at which the resistance has decreased by 10\% from the normal value and $T_c^\chi$ at which the superfluid response starts to exceed our noise floor. \textbf{b} $\chi$ extracted at $T = 0.9 T_c^\chi$. \textbf{c} $R_{\square}$ at 8 K. \textbf{d-f}, Same as \textbf{a-c} but for device B. For \textbf{e}, $\chi$ is shown at $T=0.55T_c^\chi$, for \textbf{f} $R_{\square}$ is shown at 12 K.  \textbf{g}, Inverse of the superfluid stiffness at zero temperature versus $R_{\square}$. See the main text for the method of estimation of $\rho_s(T=0)$. The colors of the circles match the colors in \textbf{a-c}. The uncertainty in the inverse superfluid stiffness from uncertainty in the height of the SQUID is indicated by the gray band. Fit to Eq.~\eqref{eq:sum_rule} shown in black. \textbf{h}, Same as \textbf{g} but for device B. }
    
    \label{fig:rvsstiff}
\end{figure*}

In Fig. \ref{fig:rvsstiff}a,d, we summarize the backgate dependence of different temperature scales characterizing the superconducting transition. We plot  $T_c^R$ at which $R_{\square}$ has decreased by 10\% from the normal state value and $T_c^{\chi}$ at which the superfluid response just rises above our noise floor. In Fig. S8 we show these temperatures $T_c$ overlaid onto the data. To compare changes in the overall strength of the diamagnetic response versus $V_\text{BG}$, we plot values of $\chi$ at a fixed fraction of $T_c^{\chi}$ for both devices. Due to the comparably low values of $T_c^{\chi}$, we plot $\chi$(.9 $T_c^{\chi}$) for device A in Fig. \ref{fig:rvsstiff}b and omit values corresponding to the two lowest values of $V_\text{BG}$. In device B, $T_c^{\chi}$ is substantially higher across the backgate voltage range. Therefore, we plot $\chi$(.55 $T_c^{\chi}$) in Fig. \ref{fig:rvsstiff}e. Lastly, we plot the normal state $R_{\square}$ in Fig. \ref{fig:rvsstiff}c,f for devices A and B, respectively.

$T_c^{R}$ and $T_c^{\chi}$ both have a non-monotonic dependence on $V_\text{BG}$ for device A and B, which is also directly visible in the data in Fig. \ref{fig:bothdev}. In device B, the superconducting transition broadens as $V_\text{BG}$ decreases, which is reflected in the growing difference between the two temperatures. In contrast to the non-monotonic dependencies of $T_c^R$ and $T_c^{\chi}$, the superfluid response increases monotonically with increasing $V_\text{BG}$ doping, whereas the normal-state resistance decreases. 

 In principle, a monotonic increase in the superfluid density and therefore $\chi$ can be expected as the normal carrier density, $n_n$, increases with $V_\text{BG}$. However, the observed change in $\chi$ over the full backgate range is larger than can be explained by a change of $n_n$ alone, especially in device B. We estimate the carrier density induced by the backgate as \(7.0 \times 10^{10}\) cm\(^{-2}\) per volt given the 300 nm thickness of the SiO$_2$ with an approximate dielectric constant of 3.8. Our setup is restricted to low magnetic fields and therefore does not allow us to perform accurate measurements of the Hall effect. To estimate the carrier density induced by the ionic gate, we compare to Refs. \cite{Iwasa2012Mos2} and \cite{Costanzo2018}, which establish a relationship between $T_c$ and $n_n$ that is consistent across many devices. Based on $T^R_c$ at $V_\text{BG}=0$, device B has a normal carrier density of 1-2 $10^{14}$ cm$^{-2}$ \cite{Costanzo2018}. Device A has a lower critical temperature, which could be due to under- or over-doping. At $V_\text{BG} < 0$, $T_c^R$ increases by approximately 0.4 K for a change of 10$^{12}$ cm$^{-2}$ in the carrier density, which is in agreement with the increase in $T_c$ observed in Refs. \cite{Chen2018,Iwasa2012Mos2,Costanzo2018} on the underdoped side of the dome. We, therefore, assume in the following that device A is underdoped with a carrier density of 0.6--0.8 x $10^{14}$ cm$^{-2}$. Compared to these carrier densities, $n_n$ decreases by at most 17\% and 11\%, in device A and device B respectively, across the accessible backgate range. However, we observe larger decreases of 25\% and 76\%, respectively, in the magnitude of $\chi$.  Similarly, the normal-state resistance of both devices changes more significantly than can be explained by the carrier density. This indicates that the electron mobility is varying with $V_\text{BG}$. From $R_{\square}$, we  estimate the mobility at $V_\text{BG} = 0$ to be 300 cm$^2$/V\,s in device A and 40 cm$^2$/V\,s in device B assuming $0.7 \times 10^{14}$ cm$^{-2}$ and $1.5 \times 10^{14}$ cm$^{-2}$ for the carrier densities respectively. These mobilities are within the range observed in the literature.

 Disorder in 2D superconductors can strongly reduce the superfluid density compared to the normal-state carrier density even at $T=0$. In the dirty limit, in which the elastic scattering rate $1/\tau$ exceeds the superconducting gap, $\Delta$, the fraction of carriers forming the superconducting condensate is expected to be $n_s(T=0)/n_n \approx 2\Delta/(\hbar/\tau)$ \cite{tinkham2004}, where $n_s(T = 0)$ is the superfluid density at zero temperature. Using $R_{\square}=m^*/(n_ne^2\tau)$, we can relate the superfluid stiffness and the sheet resistance through 
 \begin{equation}
     \rho_s(T = 0) \approx \frac{\Delta \hbar}{2 k_B e^2 R_{\square}}.
     \label{eq:sum_rule}
 \end{equation}
 To compare this relationship to our data, we model $\Mgeo$ to convert $\chi$ into $\rho_s$. $\Mgeo$ depends on our measurement geometry, such as the height and dimensions of the SQUID and the dimensions of the sample (see Supplementary Information for details). We then estimate $\rho_s(T=0)$ by fitting a phenomenological BCS model to the temperature dependence of $\rho_s$ \cite{Prozorov2006} as discussed in more detail below (see also Supplementary Information for details). We constrain the fits to below a fixed fraction of $T_c^\chi$. Due to the limited temperature range, we cannot perform this analysis for traces corresponding to the most negative $V_\text{BG}$ values from both devices. In Fig. \ref{fig:rvsstiff}g,h we plot $1/\rho_s(T=0)$ versus $R_{\square}$ for device A and B, respectively. The shaded gray areas  reflect several systematic uncertainties in estimating $\Mgeo$ which do not affect the relative changes of $\rho_s$ with backgate or temperature. The fit to Eq.~\eqref{eq:sum_rule} is shown in black and we extract $\Delta = 0.4 \pm 0.2$ meV and $\Delta = 2.6 \pm 1.2$ meV for device A and B respectively. For device B, this is in agreement with $\Delta =1.75$ meV extracted from  tunneling spectroscopy \cite{Costanzo2018}. Tunneling has not been performed in underdoped devices; however, $\Delta$ for device A is similar to the gap of overdoped devices of comparable $T_c$. Combined, our measurements suggest that the backgate modifies the  superfluid stiffness via both disorder and carrier density, rather than density alone. In some devices, such as device B, disorder tuning is the predominant mechanism.  Further, using the free electron mass and assuming $n_s^{2D} = 10^{14}$ cm$^{-2}$, we would expect  $1/\rho_s(T=0) =$ .005 K$^{-1}$. The much larger inverse stiffness we observe suggests that only a fraction of the electrons pair.
  
 Potential sources of electrostatic disorder in our devices are the SiO$_2$ substrate and the ionic gate. As a function of backgate voltage, the carrier density, screening properties, and the shape of the confinement potential may change. A striking feature in device A is a pronounced asymmetry of $T_c$ with respect to $V_\text{BG} = 0$. Refs. \cite{Chen2017,Chen2018} study few-layer MoS$_2$ devices dual-gated by an ionic gate and a backgate similar to ours. They report signatures of a low-density metallic layer forming at the bottom of the MoS$_2$ flake only at positive backgate voltage, whereas at negative backgate voltage, the carrier density of the top layer decreases. Such a metallic layer could modify the electron mean free path in the top layer by screening the disorder from the SiO$_2$ substrate, which is likely significant given the reported enhancement in mobility of gated MoS$_2$ devices that are placed on hBN \cite{Cui2015}. However, we observe a change in mobility and $\rho_s$ at all backgate voltages and no clear asymmetry in those properties with respect to $V_\text{BG} = 0$. The exact mechanism of the disorder tuning in these dual-gated devices remains an open question.
 
Next, we discuss the magnetic response close to the superconducting transition. Device A and B show significantly different behavior. In device A (Fig. \ref{fig:bothdev}b), we observe a sharp rise in the superfluid response near $T_c^\chi$, followed by a change to a lower slope giving rise to a kink in the curves. In device B (Fig. \ref{fig:bothdev}d) we observe a smooth rise in the superfluid response throughout the accessible temperature range. A jump in the superfluid response is expected in a 2D superconductor close to $T_c$ due to the BKT transition \cite{Kosterlitz1973,B.I.Halperin1979}. Just below the transition, the superfluid stiffness is predicted to satisfy the universal condition $\rho_s/T_{BKT} = 2/\pi$. In Fig. \ref{fig:bothdev}b,d, we include a line corresponding to $2 T/\pi$ converted to $\chi$ in black and the systematic error from the uncertainty in $\Mgeo$ in gray. For device A, the sharp rise in $\chi$ near $T_c$ ends at the lower edge of the universal condition error band. Instead of a jump, there is a kink, which is consistent with a BKT transition broadened and modified by a combination of finite-size and disorder effects \cite{Yong2013}. In device B, however, we do not observe any features that match the expected form of a BKT transition. 

  \begin{figure}[ht!]
    \includegraphics[width=\columnwidth]{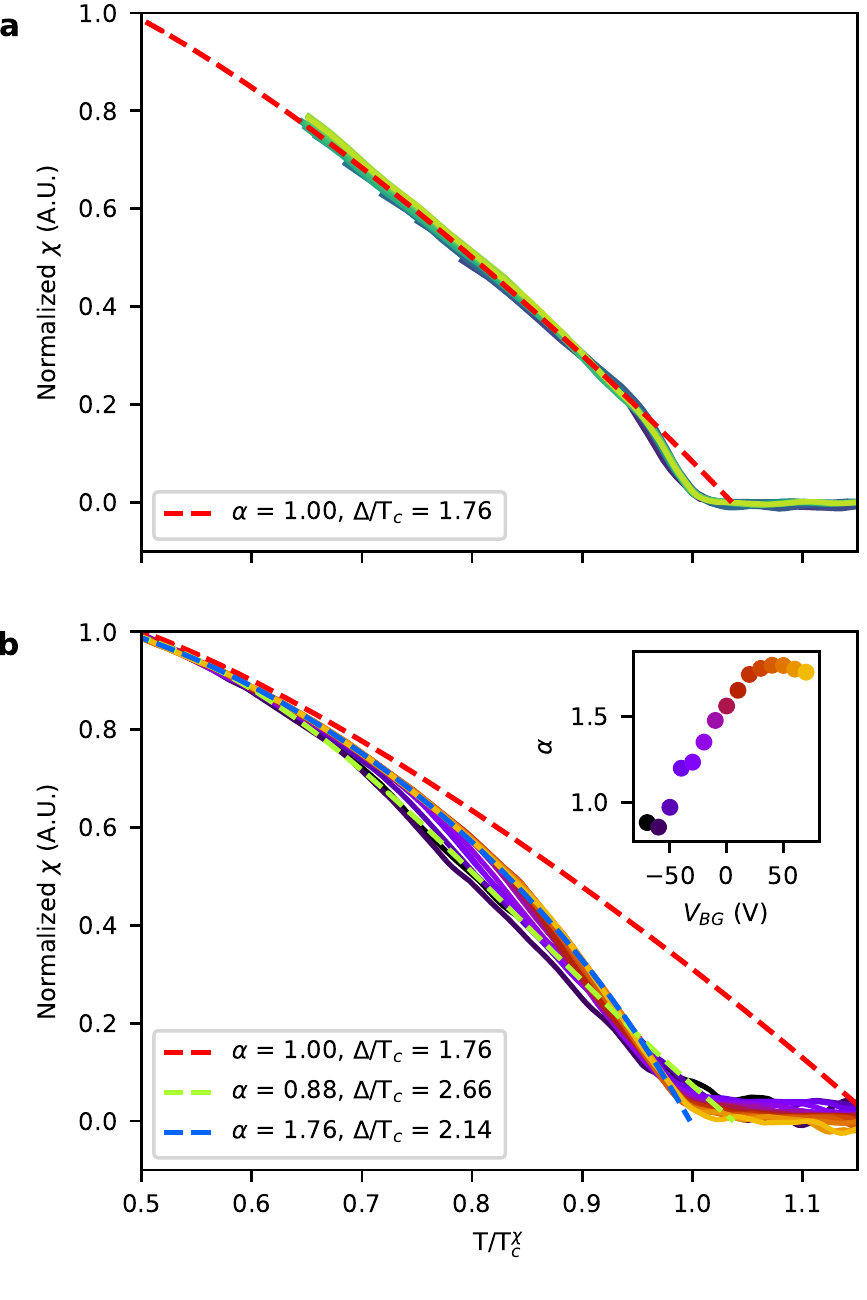}
    \centering
        \caption{\textbf{Normalized superfluid response curves. a}, Gate-tuned superfluid response \(\chi\) from device A with the temperature axis scaled by $T_c^\chi$ and the vertical axis by $\chi$ at 0.9 $T_c^\chi$ . Dashed red curve shows a fit to $\chi$ below 0.9 $T_c^\chi$ to the phenomenological model in Eq.~\eqref{eq:rho_s_model} assuming a BCS dependence with $\Delta/T_c^{BCS} = 1.76$ and a shape parameter $\alpha = 1$. \textbf{b}, Gate-tuned superfluid response of device B with the axes scaled similarly as in \textbf{a}, but using $\chi$ at .55 $T_c^\chi$. Red dashed curve shows same fit as in \textbf{a} to data below 0.55 $T_c^\chi$. The green and blue dashed curves show fits of Eq.~\eqref{eq:rho_s_model} to the full temperature range at $V_\text{BG}=-70~\mathrm{V}$ and $V_\text{BG}=70~\mathrm{V}$, respectively, with no constraints imposed on $\Delta/T_c$ and $\alpha$. The inset shows the evolution of the shape parameter, $\alpha$,  obtained from fits at all values of $V_\text{BG}$. Error bars are smaller than the markers.}
        \label{fig:collapse}
\end{figure}
To highlight how the temperature dependence of $\chi$ evolves as a function of $V_\text{BG}$, we normalize the superfluid response versus temperature curves in Fig. \ref{fig:collapse}. The vertical axis is scaled by the superfluid response at 0.9 and 0.55 $T_c^{\chi}$ for device A and B, respectively. The horizontal axis is scaled by $T_c^{\chi}$. For device A, the curves collapse. Conversely, in device B, the curves differ in the range of 0.7 to 1.0 $T_c^{\chi}$.

We compare the normalized curves to a phenomenological model for s-wave superconductors \cite{Prozorov2006}:
\begin{equation}
    \chi = \chi_0\left( 1 - \frac{1}{2T} \int_{0}^{\infty}\mathrm{ cosh}^{-2}\left(\frac{\sqrt{\epsilon^2 + \Delta^2}}{2T}\right) \mathrm{d}\epsilon\right).
    \label{eq:rho_s_model}
\end{equation}
Here $\chi_0$ is the zero temperature response, and $\Delta$ is temperature dependent gap given by:
\begin{equation}
 \begin{aligned}
    \Delta(T) = \Delta_0~ \mathrm{tanh}\left(\frac{\pi T_c^\text{BCS}}{\Delta_0}\sqrt{\alpha\left(\frac{T_c^\text{BCS}}{T} - 1\right)}\right),
 \end{aligned}
\end{equation}
where $\alpha$ is the shape parameter governing the opening of the gap, $T_c^\text{BCS}$ is the critical temperature, and $\Delta_0$ is the size of the gap at zero temperature. We first constrain $\Delta_0/T_c^\text{BCS} = 1.76$ and $\alpha = 1$ as expected for a BCS superconductor, leaving only two free parameters, $T_c^\text{BCS}$ and $\chi_0$. We fit to data below .9 $T_c^\chi$ for device A and .55 $T_c^\chi$ for device B, because the superfluid response is in clear disagreement with the phenomenological model above the cutoff temperature for device A, and in device B, the shape of the onset is changing with $V_\text{BG}$. The resulting fits for the highest backgate voltages are shown as red dashed lines in  Fig. \ref{fig:collapse}a,b. 
In device A, $T_c^\text{BCS}$ is slightly above the onset of susceptibility consistent with a small temperature range above $T_c^{\chi}$ in which the superfluid response is suppressed due to a BKT transition, which is not captured in the phenomenological model. The fitted value of $T_c^\text{BCS}$ is sensitive to details of the fitting such as an initial guess and the exact temperature range. However, the extracted low-temperature value $\chi_0$ is comparably robust and was used to estimate  $\rho_s(T=0)$ for Fig. \ref{fig:bothdev}g,h (see Supplementary Information for more detail). The onset of diamagnetism in both devices is not captured by the simple model, which could be due to a number of reasons. In particular, disorder can modify the onset of diamagnetism even in 3D superconductors. Within the phenomenological model in Eq. \ref{eq:rho_s_model}, disorder causes an increase in the shape parameter \cite{Carbotte1990,Prozorov2006}. To explore this further, we fit our data across the entire temperature range without constraints on $\alpha$ and $\Delta_0/T_c$. For device A, we cannot obtain a good fit of the model near $T_c^{\chi}$, indicating that the kink in the curves is not due to a disorder-modified BCS transition. In device B, good agreement between the model and the data can be achieved for all backgate voltages. The fits for the highest and lowest backgate voltages are shown as dashed blue and green lines, respectively, in Fig. \ref{fig:collapse}b (all fits are shown in the Supplementary Information). The fitted $\Delta_0/T_c$ significantly exceeds the BCS result, and $\alpha$ increases from less than 1 to almost 2 with increasing $V_\text{BG}$ (see inset to Fig. \ref{fig:collapse}b). This dependence is contrary to expectation, as the model suggests that $\alpha$ should decrease with decreasing disorder and hence with increasing $V_\text{BG}$, because the change of $R_{\square}$ with $V_\text{BG}$ suggests lower disorder at more positive $V_\text{BG}$.

For device A, the deviation of the superfluid response from Eq.~\eqref{eq:rho_s_model} is likely caused by phase fluctuations. That is, the kink in $\chi(T)$ results from a BKT jump in $\rho_s(T)$ slightly broadened by the interplay of finite-size and disorder effects. Although device B does not show a similarly clear feature, it is likely that similar effects are at play. For a given system size, within the disorder modified BKT paradigm we expect that stronger disorder (i.e., decreasing $V_\text{BG}$) will cause the superfluid response to become more shallow close to $T_c$ \cite{Benfatto2008}, which is what we observe. This behavior is also consistent with the generally stronger disorder in device B compared to device A as indicated by the normal state carrier mobility. 
Therefore, our data suggest that MoS$_2$ represents a crossover system, where the superfluid stiffness near $T_c$ is governed by phase fluctuations, but a clear signature of a BKT transition may or may not be present depending on doping and other parameters.

In conclusion, we report the first measurements of the superfluid response of an atomically thin van der Waals superconductor using a local probe that provides sufficient sensitivity to the small sample volume typical in this material family. We find that the superfluid stiffness monotonically increases at low temperatures as the backgate is tuned, even when the critical temperature decreases. Our analysis suggests that our devices are in the dirty limit of superconductivity in which the superfluid stiffness responds to changes in device resistivity. This demonstrates that disorder plays an important role even in crystalline 2D superconductors. Further, we observe direct signatures of a BKT transition in one device, whereas, in another the universal jump is replaced by a broad region of suppressed superfluid response close to $T_c$. This demonstrates that a clear BKT transition is not ubiquitous in these systems, but can be substantially obscured by disorder. In the present work, our 4 K base temperature prevented characterizing $\chi$ at a small fraction of $T_c$. Future work extending to lower temperatures will be sensitive to the presence of nodes in the superconducting gap, which would be a sign of an unconventional order parameter \cite{Zhou2016, Hsu2017, Liu2017, Costanzo2018}. 

\section*{Acknowledgment}
We thank Eric Smith for advice on cryogenic instrumentation and Debdeep Jena, Huili Xing, Guen Prawiroatmodjo, Eun-Ah Kim, Yi-Ting Hsu, and Debanjan Chowdhury for fruitful discussions. This work was primarily supported by the National Science Foundation under Grant No. DMR-2004864, and partially supported by the Cornell Center for Materials Research with funding from the NSF MRSEC program (DMR-1719875). This work was performed in part at the Cornell NanoScale Facility, an NNCI member supported by NSF Grant NNCI-2025233.

\section*{Author Contributions}
A.J. and K.C.N. designed the experiments. A.J., B.T.S., and G.M.F. built the instrument. A.J. fabricated the devices, performed the measurements, processed the data, and performed modeling to estimate the SQUID height and relate stiffness and magnetic response. M.L. advised on implementing liquid gating and cryogenic measurements. A.J., Y.L.L.,  N.T., and K.C.N. analyzed the data. A.J. and K.C.N. wrote the manuscript with input from all authors.

\printbibliography

\end{document}


\makeatletter 
\renewcommand{\thefigure}{S\@arabic\c@figure}
\makeatother

\title{Supplementary Information for \textit{Superfluid response of an atomically thin, gate-tuned van der Waals superconductor}}
\author{Alexander Jarjour, G.M. Ferguson, Brian T. Schaefer, Menyoung Lee, \\ Yen Lee Loh, Nandini Trivedi, Katja C. Nowack}
\date{}

\maketitle
\section{Methods}
\subsection{Device Fabrication}

\begin{figure*}[t!]
    \centering
    \includegraphics[width=\textwidth]{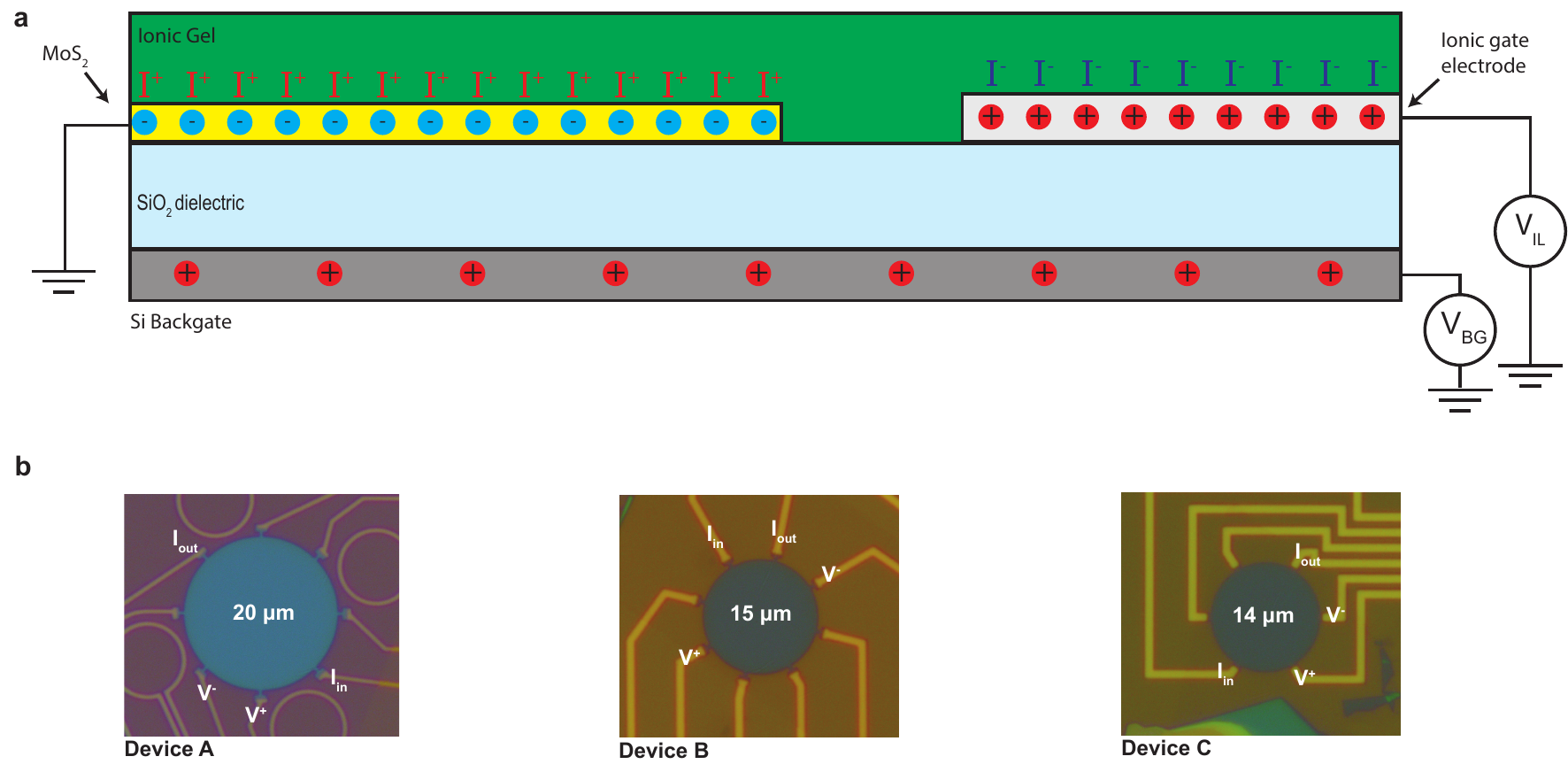}
    \caption{\textbf{Geometry of device and charge distribution. a,} Vertical cross section of a device. Thicknesses and in-plane dimensions are not to scale. The ionic gate electrode is hundreds of microns away from the MoS$_2$ device. Ions are shown as I$^+$ and I$^-$, charge accumulation is shown by colored circles. The backgate is shown at positive voltage, inducing additional electron charge accumulation in the MoS$_2$ layer. \textbf{b}, Optical images of devices A, B, C labels showing the geometry of the four point measurement used to determine the resistance. Diameter of each disk indicated at the center. Device A and B are from the main text, device C is a supplementary device discussed later in this Supplement. }
    \label{fig:charge}
\end{figure*}

We exfoliated bulk MoS\(_2\) crystals (HQ Graphene) onto flexible polydimethylsiloxane (PDMS) substrates (Dow Sylgard 184), and identified 3-10 layer flakes by optical contrast. Using a polycarbonate film supported by a PDMS stamp, flakes were individually transferred onto pre-patterned SiO\(_2\)/Si substrates. The polycarbonate film was then stripped using chloroform. The silicon substrate is highly doped to both thermalize the sample and act as a backgate, and is shown in Fig. \ref{fig:charge} in dark gray. The pre-patterned substrates are nearly entirely covered with evaporated platinum used as the ionic gate electrode (see Fig. \ref{fig:charge}) except for a clear area for the device in the center, and a thin strip in which connections run from the device to electrical bond pads. To keep leakage between this large platinum pad and the backgate low, we use Si wafers with high-quality dry chlorinated oxide (Nova Wafers), shown in light blue in Fig. \ref{fig:charge}. After transfer of the MoS\(_2\) flake, Ti/Au contacts are fabricated by a bilayer PMMA/MMA electron beam lithography lift-off process. In all e-beam steps, cold 3:1 IPA:DI developer is used. To prevent delamination of the metal, we used a mild remote oxygen plasma to descum the contact areas before deposition of the metal. To minimize contact resistance, only 1 nm of Ti is used for adhesion, achieving a contact resistance on the order of 5 kOhms/\(\mu \)m with the device doped into the superconducting state. After liftoff in acetone, we used electron beam lithography to pattern a single-layer PMMA etch mask. Using a dry CHF\(_3\)/O\(_2\) etch, the MoS\(_2\) was etched into the final device shape. The device is then baked at 250 $^{\circ}$C for 4 hours in UHV ($<$ 1e-6 Torr), to remove any absorbed water. An ionic gel is prepared by mixing 1\% polystyrene-poly(methyl methacrylate)-polystyrene triblock polymer (Polymer Source), 10\% Diethylmethyl(2-methoxyethyl)ammonium bis(trifluoromethylsulfonyl)imide (Sigma-Aldrich), and balance ethyl propionate. Ethyl propionate was chosen due to its low volatility, ability to dissolve the polymer, and miscibility with the ionic liquid. The ingredients are mixed in a dry nitrogen atmosphere in an all-glass container, and agitated for 24 hrs. Then, under a dry nitrogen atmosphere, the ionic gel is spin-coated at 2000 RPM onto the sample substrate (shown in green in Fig. \ref{fig:charge}). Without exposure to air, the solvent is removed under vacuum at room temperature for 24 hrs. After this cure, the ionic gel is manually removed over the bond pads and at the edge of the chip, preventing electrochemical reactions with the aluminum bond wire. Upon loading into the cryostat some exposure to air is inevitable, so the sample is allowed to dehydrate at room temperature in high vacuum in the system for 24 hours prior to cool down. Prior to this dehydration, all connections to the device are kept grounded. After this step, the cryocooler is turned on and allowed to cool to 4 K over 16 hours. When the sample temperature reaches 220 K, a DC voltage is applied between the ionic gate electrode and the device, causing negative ions to accumulate on the ionic gate electrode and positive ions to accumulate on the device (shown in Fig. \ref{fig:charge} as I$^-$ and I$^+$, respectively).  A compensating electronic charge appears in the MoS$_2$ device, inducing on the order of 10$^{14}$ cm$^{-2}$ charge carriers \cite{Iwasa2012Mos2}. We find that typically a voltage of approximately 5.5 V induces superconductivity, although this is likely sensitive to choices made in fabrication and design. We used 5.6 V, 5.5 V , and 5.5 V applied to the ionic gate electrode for devices A, B, and C respectively.

\subsection{Experimental Setup}

In our scanning probe microscope, we use commercial piezo stages (attocube) for coarse positioning and custom piezoelectric benders for fine positioning of the SQUID. The sample is anchored to a thermally isolated copper mount, along with a heater and a thermometer. This enables the system, and SQUID, to remain at 4 K while the sample temperature is varied. The on-chip sample leads are thermalized to the copper mount through the large area of the bond pads (60,000 \(\mu \mathrm{m}^2\)) and the high thermal conductivity of SiO\(_2\). The sample wires are thermally anchored at the baseplate of the cryostat (Montana workstation) using bobbins secured with stycast, and filtered using a QDevil RC filter. At the room temperature breakout box, unused device leads are capped using non-shorting BNC caps. Four point measurements were acquired by flowing a current of 64nA in device A and 571nA in device B. The exact four point geometry varied between devices, as shown in Fig. \ref{fig:charge}b. We used COMSOL Multiphysics to relate the measured resistance ($R_{4pt}$) to the sheet resistance we report ($R_\square$), finding $R_\square/R_{4pnt}$ = 3.6, 3.6, and 2.9 for device A, B, and C, respectively. All voltage measurements are taken with a Stanford Research SR560 preamplifier with a 100 MOhm input impedance. For the SQUID measurements, the field coil was driven at approximately 1 kHz, and the field coil current was set low enough that the sample response did not vary with the current (250 $\mathrm{\mu A_{RMS}}$ for device A, 360 $\mathrm{\mu A_{RMS}}$ for device B). The SQUID is gradiometric and therefore rejects uniform background fields. Further, the field coil is counter-wound around the two pickup loops, minimizing the mutual inductance between the SQUID and field coil when no sample is near the front pickup loop. The signal from the SQUID is amplified cryogenically by a SQUID array amplifier, and room temperature feedback electronics keeps the flux in the SQUID fixed by changing the current flowing through on-chip modulation coils. This feedback current is proportional to the flux through the SQUID, and we demodulate it using a lock-in amplifier at the frequency of the field coil drive. An air core, resistive magnet surrounds the cryostat and was used to zero the out-of-plane background field in all the measurements reported in the main text.

\subsection{Data acquisition}
We found that a common failure mode of the experiments was delamination of the ionic liquid at low temperatures causing catastrophic damage to the device. These events were often correlated with temperature changes at a finite backgate voltage. Therefore, we adopted the following procedure for our measurement: slow warming and cooling rates of less than 0.5 K/minute when the SQUID was approached to the sample, and the backgate was kept at zero when the temperature changed. Therefore for the backgate-dependent measurements, the data were acquired with temperature as the outermost sweep: for each temperature, the sample is heated to the chosen temperature, then the backgate is swept, then returned to zero, and then the temperature is changed to the next value. This is the origin of the repeated noise features line to line in some data sets.

The SQUID signal exhibits a small non-zero phase with respect to the field coil drive even far away from a sample. This is likely due to a small parasitic impedance in the electronics and wiring. This effect is negligible on the in-phase response from a superconducting sample. However, its effect on the out-of-phase response can be substantial, because that part is small or zero. Therefore, we characterized the phase shift with the SQUID far away from the sample at all frequencies and rotated the collected data to correct for this effect. After this rotation, no out-of-phase response was observed in the data reported in this work. 

\begin{figure}[t]
    \centering
    \includegraphics[width=\linewidth]{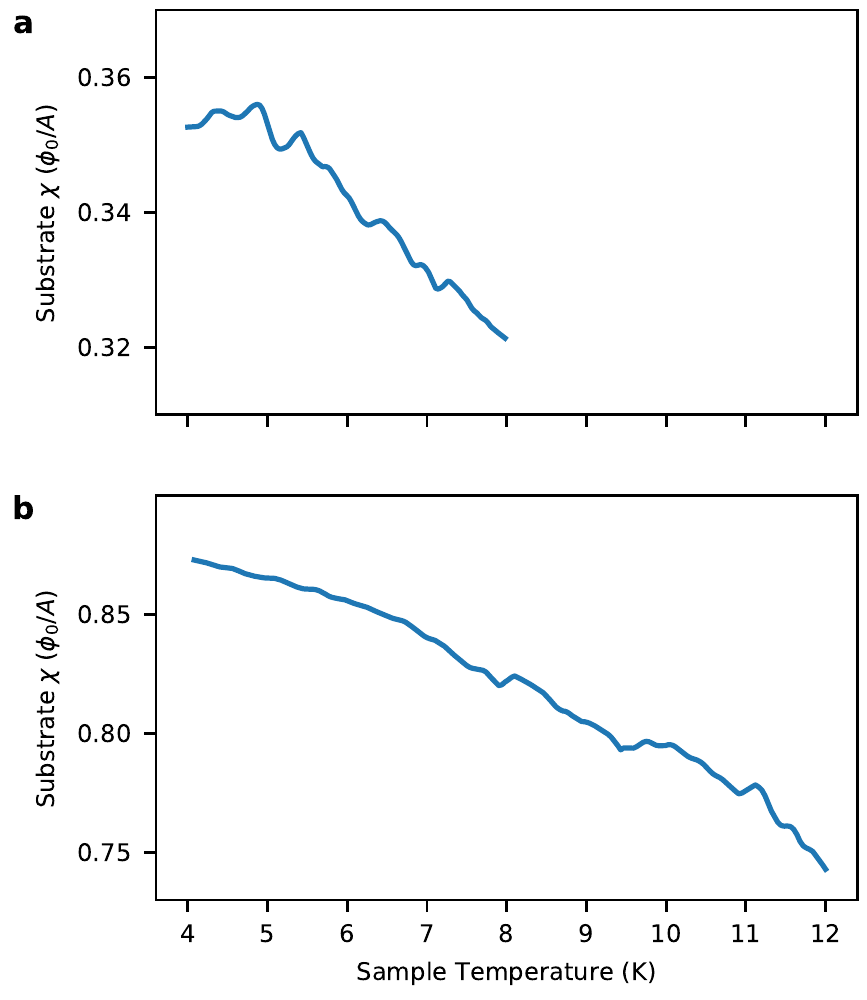}
    \caption{\textbf{Magnetic response of ionic liquid. a}, Background subtracted from data for device A. \textbf{b},
    Background subtracted from data for device B.}
    \label{fig:bgs}
\end{figure}

Two background signals contribute to our measurement signal which we subtract from the raw data. First, the field coil is counter-wound around the gradiometric squid to minimize the mutual inductance between the SQUID and the field coil. However, a slight remaining mutual inductance caused by lithographic imperfections is present such that we detect a non-zero offset signal even far above the sample. Second, close to the sample surface but away from the device, we detected a moderately temperature-dependent magnetic response likely originating from the ionic liquid. Therefore, for all measurements, we collected  a temperature-dependent background signal nearby the superconducting disk at the same height. These background data are plotted in Fig. \ref{fig:bgs} and include the offset in mutual inductance. We subtract this background from the raw magnetic response data. After that, a small amount of smoothing was applied to each resulting $\chi(T)$ curve. The data was sampled more finely in temperature than needed given the width of observed features, and therefore the smoothing only reduces the noise on the traces without changing any features. A three pass local regression algorithm was used for smoothing, with 5\% of the total data used to smooth at each point. 

\section{Additional data}

\begin{figure*}
    \centering
    \includegraphics[width=\textwidth]{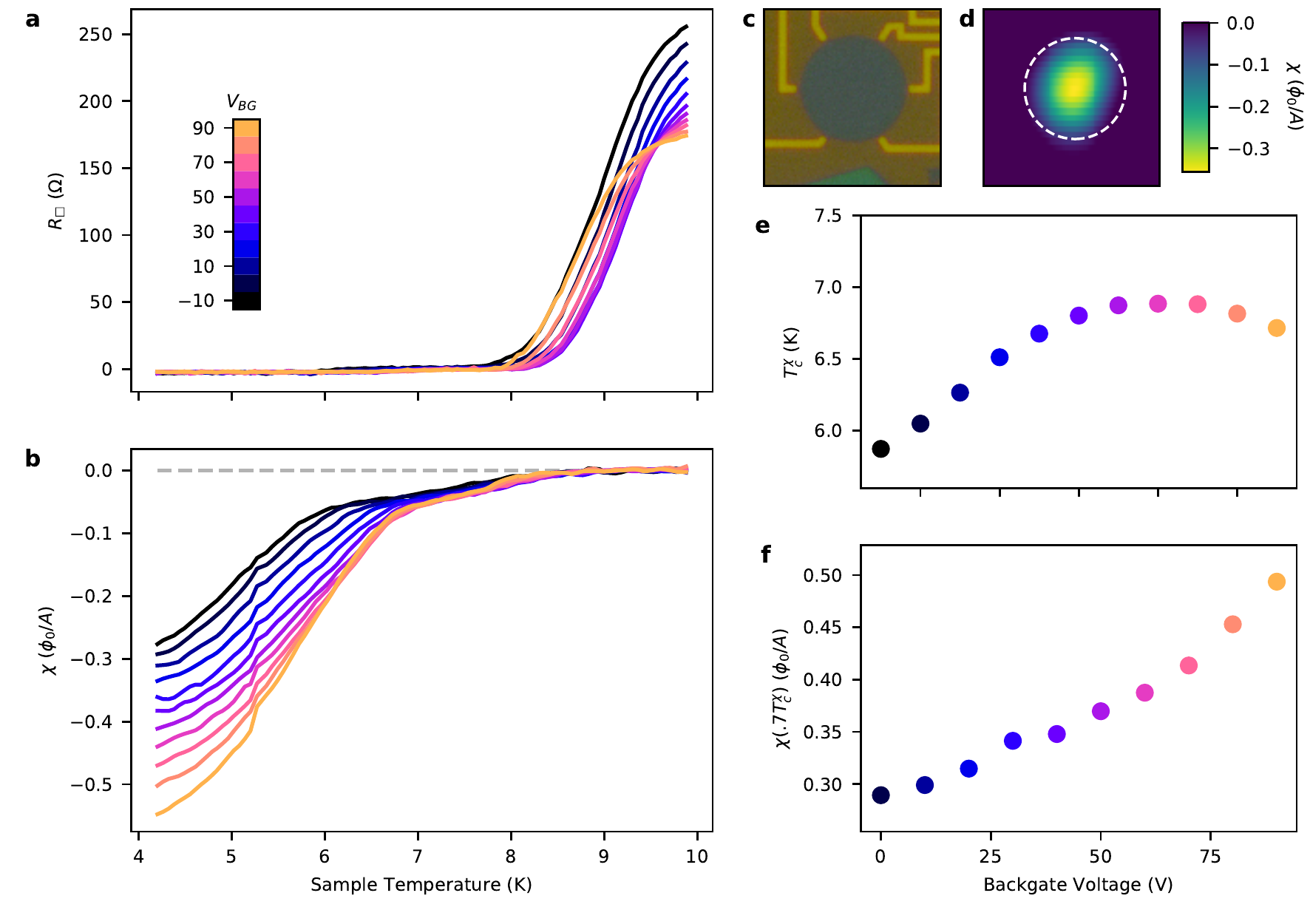}
    \caption{\textbf{Data from additional device C. a}, Sheet resistance of device C versus temperature. Different curves correspond to different backgate voltages, as indicated by the inset color bar. \textbf{b}, Magnetic susceptibility of device C. Note the initial, slow rise in diamagnetism beginning around 8 K, and then rapid increase below 6 K. Line colors correspond to the color bar in panel a. \textbf{c}, Optical image of 14.2 \(\mu\)m MoS\(_2\) disk before application of ionic gel. \textbf{d}, 4 Kelvin susceptibility map of sample. White dashed outline corresponds to 14.2 \(\mu\)m diameter of superconducting disk. \textbf{e}, Temperature of strong upturn in diamagnetism, $T_c^\chi$, versus backgate. \textbf{f}, Strength of magnetic susceptibility at 70\% of $T_c^\chi$.}
    \label{fig:devC}
\end{figure*}

In Fig. \ref{fig:devC}, we present data on an additional device labeled device C, with a diameter of 14 $\mu$m. We plot the sheet resistance versus temperature and backgate voltage. 
Some of the qualitative behavior of device A and B is also observed in device C: the low temperature superfluid response and the sheet resistivity monotonically increase and decrease with increasing backgate voltage, respectively, whereas $T_c$ changes non-monotonically. The sheet resistance is comparable to device A. For device C, we do not have on-chip structures that let us estimate the height of the SQUID (see next section), and the ionic gel was prepared in a different batch than device A and B. A quantitative analysis of the superfluid response is therefore difficult, because the SQUID height cannot be accurately estimated. In addition, the temperature range does not include sufficiently high temperatures to determine the normal state resistivity. We, therefore, omit the comparison of the low temperature superfluid stiffness and the resistivity.

The behavior of the superfluid response in device C close to $T_c$ is distinct from device A and B. In device C the superfluid response initially rises with a low slope, and starts rising sharply at a lower temperature. The onset of the superfluid response does coincide with the point at which the resistivity saturates to a low value. A possible explanation is that this behavior is related to the BKT transition, in the sense that free vortices and antivortices suppress the superfluid response initially, and that the superfluid response rise sharply at a vortex-anti-vortex binding transition. However, other explanations such as the presence of small length scale non-uniformities may also yield similar behavior. In addition, during measurements on device C, the external magnet was not included in the experimental setup yet. Therefore, a small out-of-plane magnetic field may have been present, which we estimate to be at most 14 $\mu$T or 1 flux quanta through the device. This may also modify the response of the device close to the transition.

\section{Determining the SQUID height and geometry}
We detect the superfluid response as a change in mutual inductance between the pickup loop and field coil. The magnitude of this change depends on the superfluid stiffness of the superconducting device, on the height of the SQUID above the superconductor, and the point spread function (PSF) of the SQUID. We imaged Pearl vortices in a quasi-infinite 20 nm thick Nb film to characterize the PSF. To determine the height, our SQUID is placed on a brass cantilever which forms one side of a parallel plate capacitor. By measuring this capacitance, we can detect the touchdown of the SQUID on the sample. However, even at touchdown the height of the pickup loop above the sample is unknown due to a height offset resulting from the geometry of the SQUID chip and the unknown thickness of the frozen ionic liquid. We can estimate the geometric height from the angle between the SQUID and the sample and the distance from the front corner of the SQUID chip to the SQUID pickup loop. To estimate the thickness of the frozen ionic liquid, we image the magnetic field produced by the current in a lithographically well-defined structure. To first verify this approach, we fabricated 40 $\mu$m inner diameter, 42 $\mu$m outer diameter rings on test sample without ionic liquid. We then scanned the SQUID over the ring structure and located the geometric center of the ring. Then, we varied the height of the SQUID while recording the mutual inductance between the SQUID and ring. To produce a model to fit our data, we calculated the field of a finite width current carrying annulus, and convolved this field with the previously determined PSF. To compare this model to the data, the height of the SQUID in physical volts on the piezoelectric bender must be converted into height above the sample in meters. We introduce three parameters: the offset height $z_0$ of the SQUID when we detect touchdown, the derivative of height with respect to bender voltage \textit{m}, and an effective scaling $\alpha$ of the PSF which allows the sensitivity of the SQUID to vary slightly from 1. This last parameter corrects for slight variations of the SQUID to sample angle between samples, which changes the effective area of the SQUID. If the angle, and therefore sensitivity, of the SQUID remains unchanged between the Nb film and the ring sample, we expect $\alpha$ = 1. In the case of the current carrying ring on the test sample, we fix $z_0$ by geometry and fit \textit{m} and $\alpha$. We find $\alpha = .99 \pm .02$ where the uncertainty is the fitting uncertainty. There are two errors here, the fitting uncertainty (2\%) and the difference in sensitivity from 1 (1\%). We generously estimate our overall error as 5\% in the signal magnitude. In the case of device B, we fabricated identical rings and performed the same measurements as on the test sample. We fix \textit{m} and $\alpha$ to the values found from the test sample, and fit $z_0$. We propagate the error in $\alpha$ by numerically computing derivative of the fit $z_0$ with respect to $\alpha$. We finally obtain that the height of the SQUID in device B was $2.5 \pm 1.0$ $\mu$m. This may be compared to the geometric height of the SQUID if the front corner of the chip is in firm contact, which we estimate to be 0.8 $\mu$m. Thus, we find that the ionic liquid is $1.7 \pm 1.0$ microns thick. Device A, which was measured before device B, also had current-carrying rings, but they were too small to allow meaningful measurements. Therefore, since both devices were spin coated with ionic liquid in identical processes, and the SQUID was approached using the same procedure, we assume the SQUID height was the same in both measurements. Conversely, device C was measured earlier, with a different batch of ionic liquid, and therefore may have a different height.

\section{Modeling $\Mgeo$}

To relate $\rho_s$ and $\chi$ measured at a fixed height, we need to model the screening currents flowing in the sample in response to the field applied by the field coil. From this current density, we can calculate the magnetic field produced by the sample and the amount of flux coupled into the SQUID pickup loop. To model the screening currents, we adapt a numerical model employed in the literature for two-coil measurements of superconductors to our SQUID and field coil geometry \cite{Turneaure1998}. The disk-shaped sample is modeled as a finite number $N$ of infinitely thin superconducting rings with the radius of the \(i^{th}\) ring given by \(\mathbb{R}_i\) reflecting the circular symmetry of the problem. We approximate the field coil as circularly symmetric generating a magnetic field in the sample plane that is determined by a vector potential \(\mathbb{A}(t) = \mathbb{A}_0 e^{i\omega t}\). Here the boldface indicates a vector indexing over the rings. The vector potential may vary in the radial and out-of-plane directions. Using Ohm's law with a complex conductivity, the current in the superconducting disk can be written as:

\begin{equation}
    \mathbb{J}\ = - (\sigma^{2D}_1 + i\sigma^{2D}_2)i \omega\left(\frac{1}{2 \pi \mathbb{R}} M w \mathbb{J} + \mathbb{A}_0\right)
\end{equation}

 \(\mathbb{J}\) is a vector indexing over the rings with \(\mathbb{J}_i\) the current in \(i^{th}\) ring. \(M_{i,j}\) is the mutual inductance between the \(i^{th}\) and \(j^{th}\) rings. The term \(M\mathbb{J}\) takes the field created by the superconductor itself into account. The division by \(\mathbb{R}\) is performed element-wise onto the result of \(M\mathbb{J}\). Typically in a superconductor, $\sigma^{2D}_1 = 0$ and \(\sigma_2^{2D} = -2/(\omega \mu_0 \Lambda) = - 4 k_B e^2 \rho_s/(\hbar^2\omega)\). Because \(\mathbb{A}_0\) is the applied vector potential, \(\mathbb{J}\) can be computed in principle by a matrix inversion. We compute the coefficients of \textit{M} for $i\neq j$ by using a closed form in terms of elliptic functions for the vector potential of a current carrying ring \cite{Garrett1963}. For \( M_{i,i} \) we use the results from Ref. \cite{Gilchrist1996}. To calculate \(\mathbb{A}_0\), the field coil is modeled as carrying a circular current with a finite radial width and zero thickness in the z-direction. Then, the magnetic field everywhere can be computed using the resulting \(\mathbb{J}\). Finally, we convolve the PSF discussed in the previous section with this magnetic field to compute the flux through the SQUID. We repeat this calculation as a function of $\rho_s$ to create a conversion table between a measured value of $\chi$ and the corresponding value of $\rho_s$. To estimate the uncertainty coming from uncertainty in the SQUID height, we also perform this calculation for different heights, and apply the uncertainty in the height previously found to predict the uncertainty in stiffness reported in the main text. 

\begin{figure}
    \centering
    \includegraphics[width=\linewidth]{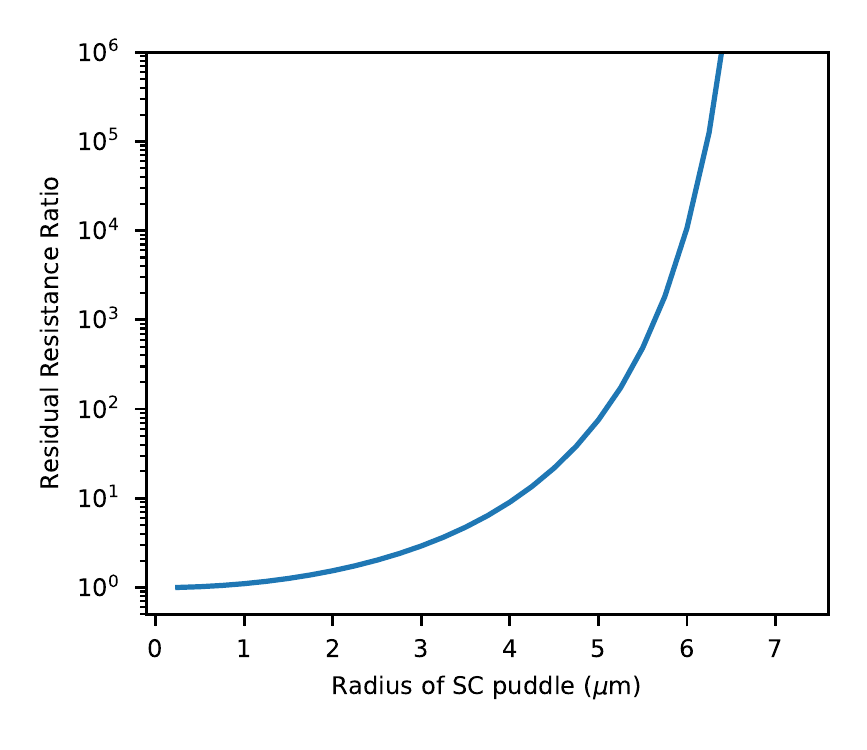}
    \caption{\textbf{Simulation of the residual resistance ratio.} Ratio of the normal resistance of the device to the observed resistance in the numerical model, as a function of the radius of the superconducting region in the center.}
    \label{fig:radiusdependence}
\end{figure}

\begin{figure*}
    \centering
    \includegraphics[width=\textwidth]{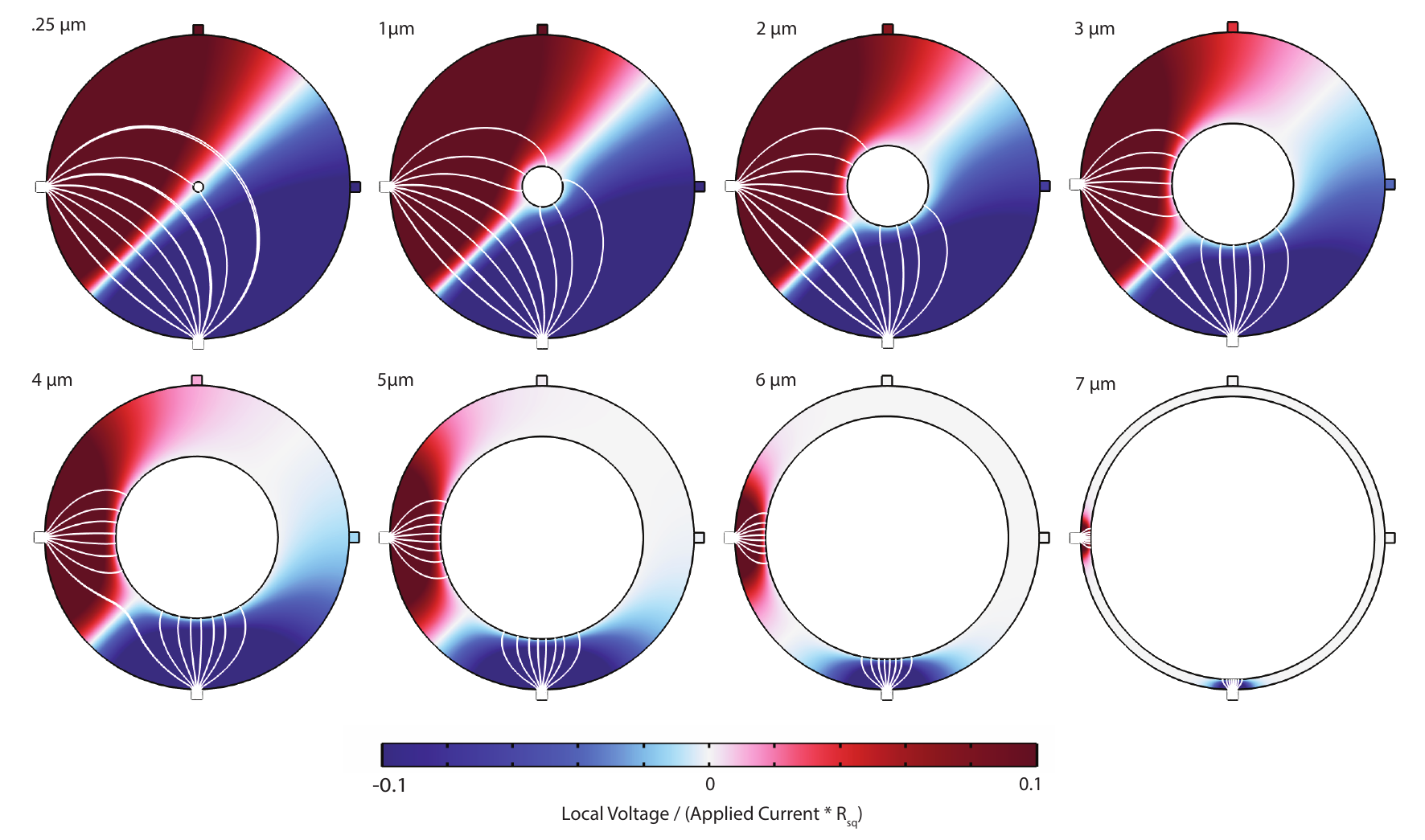}
    \caption{\textbf{Simulation of voltage in normal metal ring/superconductor device.} Current is not simulated in superconducting region. Disks demarcated by radius of superconducting region. All disks have 7.5 \(\mu m\) outer radius. Color scale is dimensionless, local voltage per unit current applied and unit sheet resistance. White streamlines show current density}
    \label{fig:normalmetalrings}
\end{figure*}

\section{Effect of a normal conducting region along the circumference of a disk-shaped device}
Our devices exhibit a backgate-dependent finite resistance even when we detect a superfluid response. We do not detect any obvious non-uniformities in the superfluid response that can explain this finite resistance. Here we explore if the presence of a normal conducting ring around the perimeter of the disk-shaped devices can explain the observed finite resistance. Such a ring may form due to a combination of a non-uniform carrier density profile and the dependence of the critical temperature (and presence of superconductivity) on density. A non-uniform density profile may be present in our devices for several reasons: The backgate fringing fields can cause non-uniform density; this effect is confined to a few oxide thicknesses (300 nm) away from the edge of the device. In addition, ionic liquid gating of our devices is a complicated process, especially since the gate is set near the freezing temperature of the ionic liquid. If the accumulation dynamics cause a non-uniform distribution of ions above the device, in particular a lower density of ions close to the device boundary, this would give rise to significantly lower carrier density closer to the edge of the device. 

\begin{figure}
    \centering
    \includegraphics[width=\linewidth]{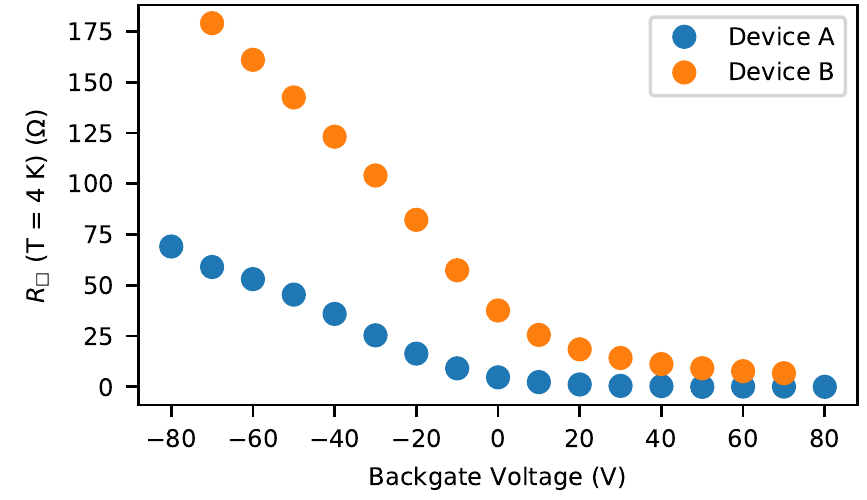}
    \caption{\textbf{Residual Resistance.} Residual resistance in the superconducting state in device A and B. Note, this is the $R_{\square}$ resistance, not the measured four-point resistance, chosen so that the resistance values are consistent with the main text. }
    \label{fig:res_vs_bg}
\end{figure}

In the following, we carry out simulations to estimate how wide a normal conducting ring is needed to explain the observed finite resistance. This width does depend on the assumed resistivity.  We solve the following model system: a superconducting disk of radius \(R_{SC}\) inside a normal metal ring of outside radius \(R\). Electrical contact is made to the disk at 4 points, equally spaced around the outside radius \(R\). Two adjacent contacts are used as the source and drain contacts, and the remaining two contacts are used as the voltage probes. We assume a conductor with zero resistance for the superconducting region, but ignore effects from the magnetic fields produced by the current density,  which should be negligible given the long Pearl length of the superconducting samples. By symmetry, the superconductor must be at fixed potential halfway between that of the source and drain. Further, by virtue of the infinite conductivity of the superconductor, it must have this potential everywhere. Therefore, we can find the local potential in the normal conducting, annular region by applying a potential between the source and drain contacts and treating the boundary with the superconducting region as an equipotential line. This approach avoids instabilities associated with the large conductivity of the superconducting region when solving for the local potential numerically. Using COMSOL multi-physics, we then solve for the current flowing through the ring, and compute the corresponding resistance of the device measured in the four-point geometry. The voltage at which the superconductor is held is checked by ensuring that no net current flows in or out of the superconducting region. Figure \ref{fig:normalmetalrings} shows simulations in which the current is applied at the left contact, and drained at the bottom contact for different widths of the normal conducting region. The local potential is proportional to the applied current and the sheet resistance of the film. Therefore, we plot the local potential voltage divided by the product of the applied current and the sheet resistance, which is dimensionless. The color scale is chosen to highlight the potential near the voltage probes at the top and right sides of the device. From the simulations, we can extract the ratio of the resistance measured when the full disk is normal to the resistance measured when the circular region is superconducting, $RRR_{SC}$. This ratio is shown in Figure \ref{fig:radiusdependence} as a function of the radius of the superconducting region while assuming a total diameter of 7.5 \(\mu\)m for the disk. When the radius of the superconducting region is reduced to .25 \(\mu\)m, the ratio becomes 1, as the device is nearly entirely non-superconducting. Conversely, as the radius approaches 7.5 \(\mu\)m, the lithographic radius of the disk, the ratio diverges. In our devices, we find ratios of the normal state resistance to resistance below the superconducting transition as low as approximately 5. This would require an unrealistically large width of the normal ring, if its resistivity is assumed to be equal to the resistivity suggested by the normal state resistance. However, the resistivity close to the edge may be higher. Interestingly, the residual resistance in device A and B display a pronounced asymmetry with respect to $V_{\textrm{BG}}=0$ (see Fig. \ref{fig:res_vs_bg}). In addition, device C shows almost vanishing resistance below $T_c$. 
Finally, we note that the same width normal region along the device boundary has a much more pronounced effect in disk-shaped devices than it has in a traditional Hall bar geometry. In particular, for high-aspect ratio Hall bars with voltage probes far from the source and drain, the residual resistance will be comparatively small as nearly all the current will enter the superconductor before the region where the voltage measurement is taken. Circular devices are not an ideal geometry for transport, however we utilize them to dramatically simplify the modeling of the SQUID data.

\begin{figure}
    \centering
    \includegraphics[width=\linewidth]{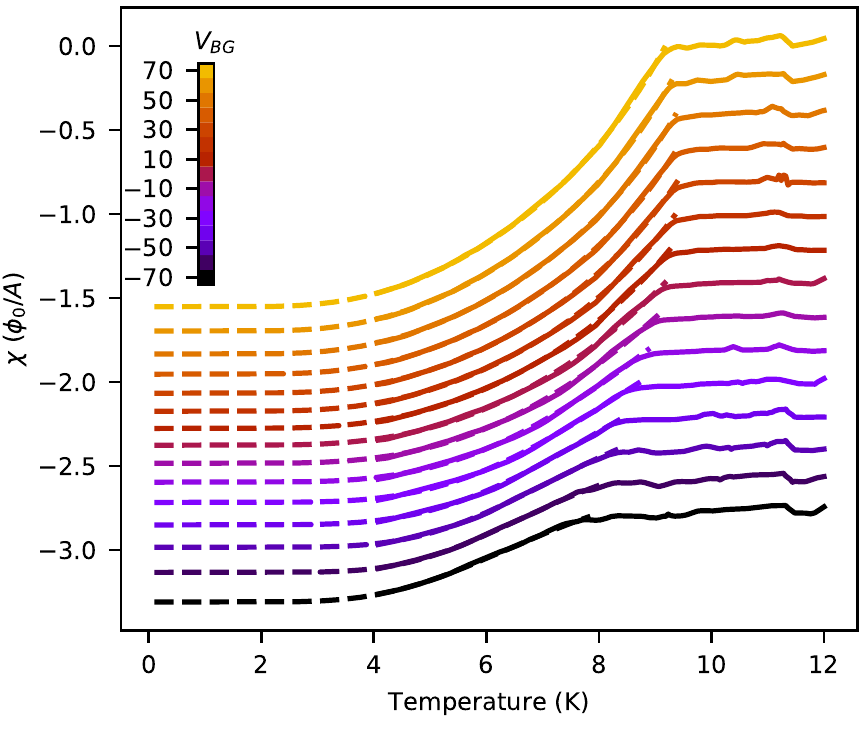}
    \caption{\textbf{Full BCS fits to Device B.} Full fittings of the phenomenological model to the data from device B. Traces are vertically offset for clarity.}
    \label{fig:bcsfitting}
\end{figure}

\section{Bounding the low temperature stiffness}

In Fig. 3g,h of the main text, we report the inverse stiffness at low temperature versus sheet resistance. We compute the low temperature stiffness by fitting the phenomenological model in Eq. 16 from Ref. \cite{Prozorov2006}, and constrain $\alpha = 1$ and $\Delta/T_c$ = 1.76. Here, we show those fits, and place additional bounds on the possible low-temperature superfluid stiffness. Generally, in both conventional and unconventional superconductors, $d\rho_s/dT$ is negative and steadily increases in absolute value as $T$ increases \cite{Prozorov2006}, i.e. as we approach lower temperatures the curve tends to flatten. Therefore, barring unusual order parameters like non-monotonic d-wave, we can place an upper bound on the magnitude of low temperature stiffness by extrapolating from 4 K to 0 K using a line with the slope of $\rho_s$ at 4 K. Further, the value of $\rho_s$(4 K) provides a lower bound for $\rho_s$(0 K). In Fig. \ref{fig:limits on stiffness}a, we show the superfluid response curves for device A extrapolating from left to right with a constant line, the 4K slope, and the fit to T = 0K. Fig. \ref{fig:limits on stiffness}b shows the same for device B. Figs. \ref{fig:limits on stiffness}c,d show $\rho_s^{-1}(T = 0)$ estimated from the three methods shown in  Fig. \ref{fig:limits on stiffness}a,b. 

\printbibliography

\begin{figure*}
    \centering
    \includegraphics[width=\textwidth]{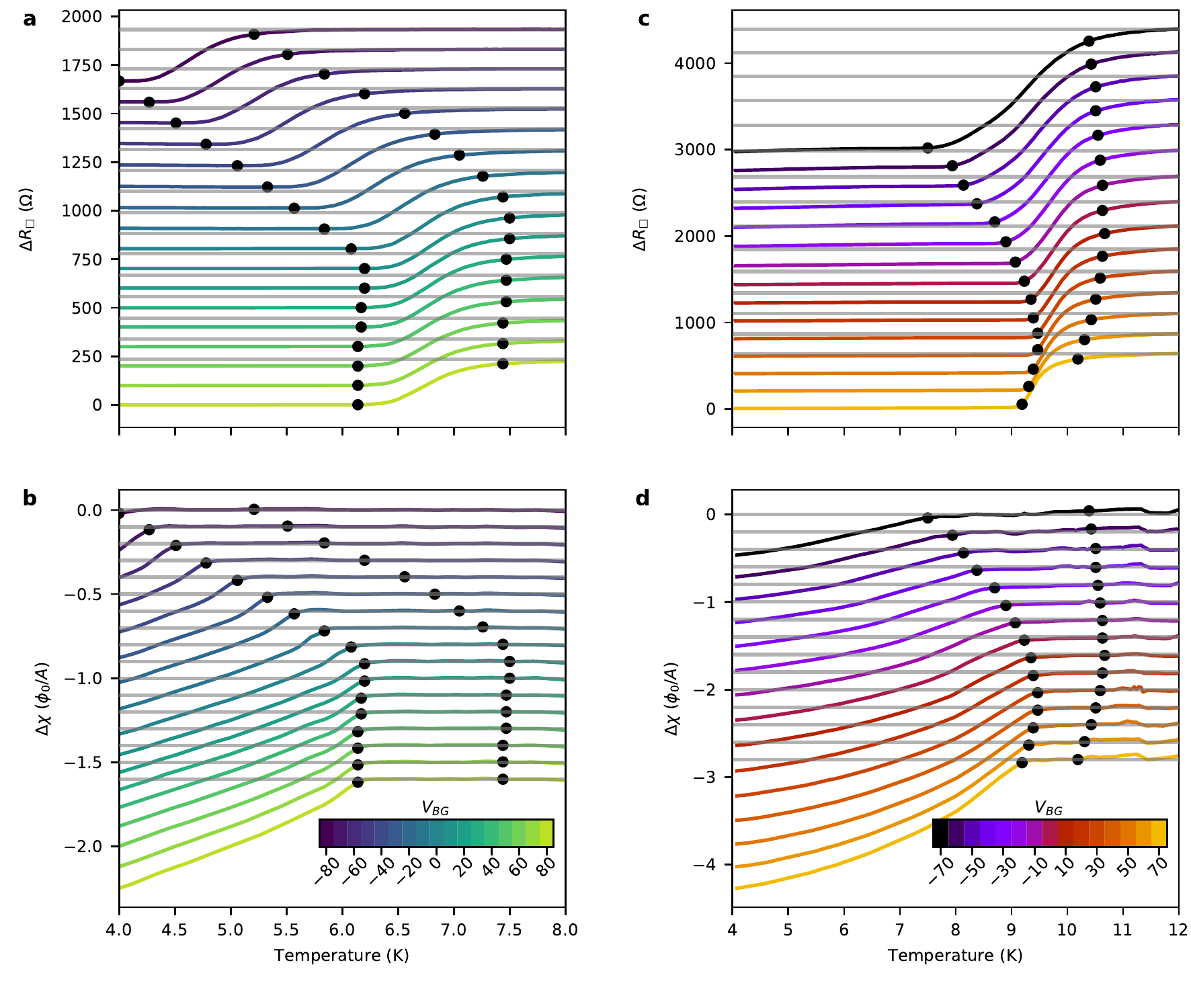}
    \caption{\textbf{$R_\square$ and $\chi$ for device A and B with $T_c^{\chi}$ and $T_c^R$ indicated. a}, Resistance data from Fig. 2a in the main text, offset in the vertical direction. Gray lines indicate 8 K resistance. Black circles at higher temperature are $T_c^R$, those at lower temperature are $T_c^\chi$. \textbf{b}, Magnetic response curves from Fig. 2b in the main text, offset in the vertical direction. Gray lines indicate $\chi = 0$. Black dots the same as in a. \textbf{c}, Same as a, but for device B. \textbf{d}, same as b, but for device B.}
    \label{fig:offset_data}
\end{figure*}

\begin{figure*}
    \centering
    \includegraphics[width=\textwidth]{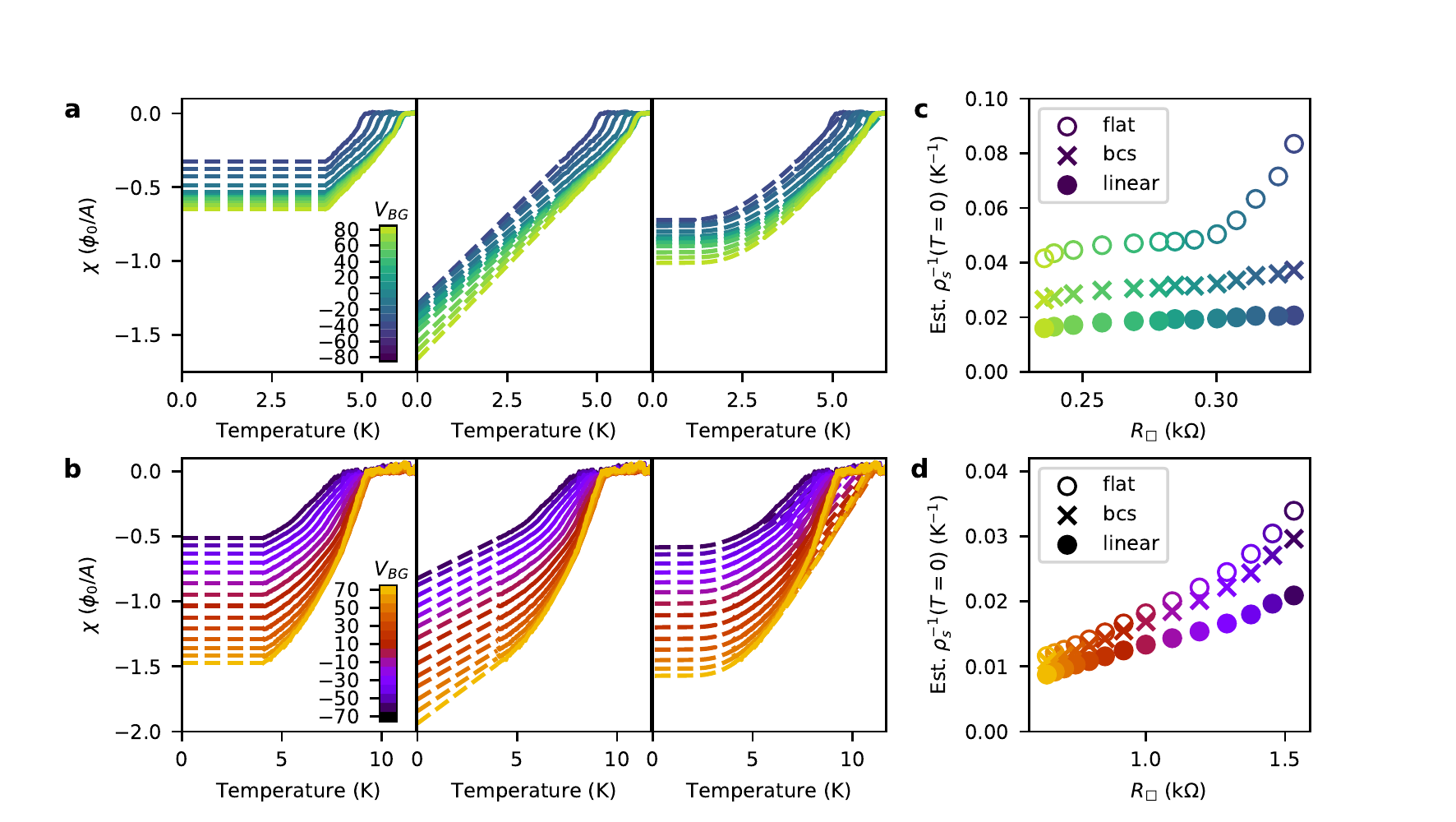}
    \caption{\textbf{Placing bounds on low temperature superfluid stiffness, $\rho_s(T=0)$. a}, Three methods of extrapolating $\chi$ in device A to zero temperature. From left to right: the flat method which assumes no more growth in $\chi$ occurs, the linear method, which linearly extrapolates from the slope at the lowest temperature, and the fitting of the BCS model with $\alpha$ = 1 and $\Delta/T_c$ = 1.76 to the data below 90\% of $T_c$ (the method used in Fig. 3c of the main text). \textbf{b} Three methods of extrapolating $\chi$ in device B to zero temperature. From left to right: the flat method which assumes no more growth in $\chi$ occurs, the linear method, which linearly extrapolates from the slope at the base temperature, and the fitting of the BCS model with $\alpha$ = 1 and $\Delta/T_c$ = 1.76 to the data below 55\% of $T_c$ (the method used in Fig. 3d of the main text).\textbf{c} Calculated device A zero temperature stiffness from the three methods. Note the poor agreement between the flat method and the other two methods for the highest resistance, most negative backgate curves. This is due to the clear proximity of $T_c$ to the base temperature in these data, leading to very little stiffness at base temperature and hence a large inverse superfluid stiffness. \textbf{d} Calculated device B zero temperature superfluid stiffness from the three methods. Note the quite good agreement between the three methods, better than the height error reported in the main text.}
    \label{fig:limits on stiffness}
\end{figure*}